\documentclass[11pt]{article}
\usepackage[a4paper,hmarginratio=1:1,vmarginratio=2:3,
totalwidth=15.6cm,totalheight=22.5cm]{geometry}
\usepackage{bm,epstopdf,epsfig,amsmath,amssymb,amsfonts,colordvi,latexsym,comment,cancel,
verbatim}
 \usepackage{graphicx,graphics}
 \usepackage[font=md,captionskip=8pt]{subfig}
 \usepackage[usenames,dvipsnames]{color}

\usepackage{setspace}
\newcommand{\si}{{\sigma}}

\newcommand{\cL}{{\cal L}}

\newcommand{\cO}{{\cal O}}

\newcommand{\ra}{\rightarrow}
\newcommand{\be}{\begin{equation}}
\newcommand{\ee}{\end{equation}}
\newcommand{\bea}{\begin{eqnarray}}
\newcommand{\eea}{\end{eqnarray}}

\newcommand{\baa}{\begin{array}}
\newcommand{\eaa}{\end{array}}
\long\def\symbolfootnote[#1]#2{\begingroup
\def\thefootnote{\fnsymbol{footnote}}\footnote[#1]{#2}\endgroup}
\begin{document}
\begin{flushright}
CERN-PH-TH-2014-080\\
\today
\end{flushright}

\thispagestyle{empty}

\vspace{2cm}

\begin{center}
{\Large {\bf Naturalness in low-scale SUSY models

\bigskip
and   ``non-linear'' MSSM}}
\\
\medskip
\vspace{1.cm}
\textbf{I. Antoniadis$^{\,a}$, E. M. Babalic$^{\,b}$, D. M. Ghilencea}$^{\,a, b,\,}$

\bigskip
\medskip
$^a$ {\small CERN Theory Division, CH-1211 Geneva 23, Switzerland}

$^b$ {\small Theoretical Physics Department, National Institute of Physics and}

{\small  Nuclear Engineering (IFIN-HH) Bucharest, MG-6 077125, 
Romania.}

\end{center}
\bigskip

\begin{abstract}
\noindent
{\small 
In MSSM models with various boundary conditions for the soft breaking 
terms ($m_{soft}$) and for a higgs mass of 126 GeV, 
there is  a (minimal) electroweak  fine-tuning $\Delta\approx 800$ to $1000$ for
the constrained MSSM and  $\Delta\approx 500$ for non-universal gaugino masses. 
These values, often regarded as unacceptably large, may indicate a problem of 
supersymmetry (SUSY) breaking, rather than of SUSY itself. A  minimal  
modification of these models is to lower the SUSY breaking scale  in the 
hidden sector ($\sqrt f$) to few TeV, which we show to
restore  naturalness to more acceptable levels  $\Delta\approx 80$ for the
most conservative case of low $\tan\beta$ and ultraviolet boundary 
conditions as in  the constrained MSSM.  This is done without 
introducing additional fields in the visible sector, 
unlike other models that attempt to reduce $\Delta$.
In the present case $\Delta$ is reduced due to additional (effective) quartic higgs 
couplings proportional to the  ratio  $m_{soft}/\sqrt f$ 
of the visible to the hidden sector SUSY breaking scales.
These couplings are generated by the auxiliary component of the goldstino superfield.
The model is discussed in the limit its sgoldstino component 
is integrated out so this superfield is realized non-linearly (hence the name of the model)
while the other MSSM superfields are  in their linear realization.
By  increasing the hidden sector scale  $\sqrt f$ one obtains a continuous 
transition for fine-tuning  values,  from  this model to the usual 
(gravity  mediated)  MSSM-like models.}
\end{abstract}

\newpage

\section{Introduction}

If supersymmetry (SUSY) is realized in Nature, it should be broken at some high scale.
A consequence of SUSY breaking is the existence of a Goldstone fermion - the goldstino -
and its scalar superpartner, the sgoldstino. The goldstino becomes the longitudinal component
of the gravitino which is rendered massive (super-Higgs mechanism), with a  mass of order
$f/M_P$ where $\sqrt f$ is the scale of spontaneous 
supersymmetry breaking in the hidden sector and $M_P$ is the Planck scale.
Also, the sgoldstino  can become massive and decouple at low energies.
One interesting possibility is that 
$\sqrt f\ll M_P$ which represents the case of so-called 
low-scale SUSY breaking models that we analyze in this work. 
Then the longitudinal gravitino
component couplings  which  are those of the goldstino and
 proportional to $1/\sqrt f$ \cite{cddfg} are much stronger 
than the couplings of the  transverse gravitino component fields which 
are Planck-scale suppressed. The latter vanish in the gravity-decoupled 
limit and one is left with  a goldstino superfield besides the matter and vector superfields
of the model. 
The gravitino is then very light, in the milli-eV range if SUSY breaking is in the 
multi-TeV region.

In this work we consider a variation of the minimal supersymmetric standard model (MSSM) called
``non-linear MSSM''
defined in \cite{nonlinearMSSM} (see also \cite{Petersson:2011in,SK}) 
in which  $\sqrt f$ is a free parameter that can be
as low as few times the scale of soft breaking terms  in the visible sector,
denoted generically  $m_{soft}$.
We assume that all fields beyond the MSSM spectrum (if any)  are heavier than $\sqrt f$
(including the sgoldstino). 
Then, at energies of few TeV, $E\!\sim\! m_{soft}\! < \!\sqrt f$ we have the MSSM fields and
the (non-linear) goldstino superfield ($X$) coupled to them. The auxiliary component
field $F_{X}$ (with $\langle F_X\rangle\sim - f$)  of $X$ 
 can mediate interactions ($\propto 1/f$)
between the MSSM fields and  generate sizeable effective couplings, 
in particular in the Higgs sector, if $\sqrt f$ is low (few TeV). 
The study of their  implications for the electroweak (EW) fine-tuning 
is one main purpose of this work. This energy regime 
can be described by a nonlinear goldstino superfield\footnote{hence the name of the model
as ``nonlinear'' MSSM.}, that satisfies $X^2\!=\!0$ \cite{SK,Rocek:1978nb,Lindstrom:1979kq}.
This constraint  decouples (integrates out) 
the scalar component of $X$ (sgoldstino), independent of the visible sector details
(it depends only on the hidden sector \cite{sg}).
The alternative case of a light sgoldstino, that can mix with the Standard Model 
(SM) higgs,  was studied in \cite{Petersson:2011in,Demidov}.
At even lower energies, below  the sparticle masses one is left with the goldstino 
fermion coupled to SM fields only, and all supermultiplets are realized nonlinearly, 
i.e. all superpartners are integrated out.

However, with so far negative searches for supersymmetry at the TeV-scale, 
the original motivation for SUSY, of solving the hierarchy problem, is sometimes
questioned, since  the stability at  the quantum level of the hierarchy
 EW scale $\ll M_P$  becomes more difficult to respect.
Indeed, the EW scale 
$v^2=-m^2/\lambda$ where $m$ is a combination of soft  masses ($m_{soft}$),
therefore  $m\sim$~TeV and $\lambda\sim \cO(1)$ an effective quartic higgs coupling;
 with an increasing $m\sim m_{soft}$, it is more difficult to obtain $v=246$ GeV. 
This tension is quantified by EW scale fine-tuning measures hereafter denoted generically
$\Delta$ with two examples  $\Delta_m$,  $\Delta_q$   \cite{ft,Anderson:1994dz}
(early studies in \cite{Barbieri:1998uv}) defined as
\medskip
\bea\label{ss1}
\Delta_m=\max\big\vert \Delta_{\gamma^2}\big\vert,
\quad
\Delta_q=\Big\{\sum_{\gamma} \Delta_{\gamma^2}^2\Big\}^{1/2}\!\!\!,
\,\,\,\,\, {\rm with}\,\,\,
\Delta_{\gamma^2}\equiv \frac{\partial \ln v^2}{\partial\ln\gamma^2}, 
\eea
$\Delta_q$ and $\Delta_m$ quantify the variation of $v$ under small relative
 variations of the ultraviolet (UV) parameters $\gamma$ that denote
the SUSY breaking parameters  and the (bare) higgsino mass ($\mu_0$).
$\Delta_{m,q}$ are regarded as  intuitive 
measures of  the success of SUSY as a solution to the
hierarchy problem. For the constrained MSSM,  $\gamma$ denotes the set: $m_0$, $m_{12}$, $\mu_0$,
$A_t$, $B_0$. For the recently measured Standard Model-like 
 higgs mass $m_h\approx 126$ GeV \cite{mh},  {\it minimal}  
values of $\Delta_{m,q}$ in the constrained MSSM are $\approx 800-1000$ \cite{KR},
reduced to $\approx 500$ for non-universal boundary conditions for gauginos.
 These values are rather far from those regarded 
by theorists as  more ``acceptable'' (but still subjective)
of $10$ to $100$.

One can ask however  what  relevance such 
values of the EW fine-tuning have for the realistic character of a model
and whether less subjective,  model-independent bounds actually exist.
Recent results \cite{GG} (based on previous \cite{KR,Casas})
 suggest that there is an interesting link 
between the EW fine tuning and  the minimal value 
of chi-square ($\chi^2_{min}$) to  fit  the EW observables.
Under the condition that motivated SUSY of {\it fixing} the EW scale $v=v(\gamma)$
to its value (246 GeV) and with some simplifying assumptions it was
found  that there exists a model independent upper bound 
 $\Delta_q\ll \exp(n_{df})$ \cite{GG};
here $n_{df}$ is the number of degrees of freedom of the
model, $n_{df}=n_\cO-n_p$ with $n_\cO$ the number of observables and $n_p$ the number 
of parameters. Generically,  $n_{df}\sim 10$ or so, see for example
 Table 1 in second reference in  \cite{GG},
 depending on the  boundary conditions of the MSSM-like model. 
This gives $\Delta_q\ll \exp 5\approx 150$
or so. This is  an estimate of the magnitude 
one should seek for $\Delta$ and supports the common view mentioned above
 that a tuning  $\Delta_q\approx 100$ is ``acceptable''. It 
should be noted however, that the nearly exponential dependence of minimal 
$\Delta_{m,q}\approx \exp (m_h/{\rm GeV})$ noticed in \cite{Cassel:2010px} 
and the theoretical error of 2-3 GeV of the Higgs mass \cite{error}
bring an error factor to the ``acceptable'' value of $\Delta$ as large as
$\exp(2)\approx 7.4$ (or $\exp (3)\approx 20$). Therefore
any value of $\Delta$ should be regarded with due care. 
Nevertheless, the above results tell us that a small  $\Delta$ is preferable.

This view is further confirmed by a less conservative approach
which shows  that there is also a link between the EW fine tuning and the 
covariance matrix of a model \cite{Dreiner,Ghilencea:2013nxa} in the basis of UV parameters
($\gamma$). 
This matrix  was shown  \cite{Ghilencea:2013nxa}  to  automatically  contain
contributions due to the EW fine-tuning wrt parameters $\gamma$ and, in particular, 
the trace of its inverse contains a contribution proportional to $\Delta_q$. 
As a result, imposing  a fixed, s-standard deviation 
of the value of chi-square  $\chi^2$  of a model from its minimal 
value $\chi^2_{min}$ i.e. $\delta\chi^2\!\leq\! s^2$,  
($\chi^2\!=\!\chi^2_{min}\!+\!\delta\chi^2$) then demands in the loop order considered
that  $\Delta_q$  have an upper bound~\cite{Ghilencea:2013nxa}.
This is a model-independent result and supports
our motivation here of seeking  models with low $\Delta$.

A very large  EW fine tuning,  that  increases further 
with negative searches for SUSY may suggest that we do not understand well the mechanism 
of SUSY breaking (assuming that SUSY exists not far above the TeV-scale).
This motivated us to consider the models with low 
SUSY breaking scale mentioned above and to evaluate their EW fine-tuning.
A previous analysis in such models  can be
 found in \cite{Casas:2003jx,Brignole:2003cm}.
We examine the values of both $\Delta_m$ and $\Delta_q$ in the ``non-linear MSSM''
 \cite{nonlinearMSSM} which has a low scale of SUSY breaking, 
$\sqrt f\sim$ few TeV. The only difference of this model 
from the  usual MSSM is present in the gravitino/goldstino and dark matter sectors.
We show that this model can have a reduced fine-tuning compared to that in the MSSM-like 
models. The reduction is done without  additional parameters or extra 
fields in the ``visible'' sector which is unlike other models 
that reduce EW fine-tuning by enlarging the spectrum. 
 Our  results depend only on the ratio $m_{soft}^2/f$
 of the SUSY breaking scale in the visible  sector 
to that in the hidden sector.  When $\sqrt f$ is low
(few TeV) we are in the region of low-scale-SUSY breaking models (with light gravitino)
 while at large $\sqrt f\sim 10^{10}$ GeV we recover the MSSM-like models.
We thus have an interpolating parameter between these classes of models. The 
reason why EW fine-tuning is
reduced is due to additional quartic higgs 
interactions  mediated by the auxiliary component of the goldstino superfield,
as mentioned earlier;  these  enhance the {\it effective} higgs coupling 
$\lambda$ and even increase the higgs mass already at tree level.
We stress that this behaviour is generic to low-scale SUSY models.

In the next section we review the model. In Section~\ref{section3}
we compute analytically the one-loop corrected higgs mass including
$\cO(1/f^2)$ corrections from effective operators generated by SUSY breaking. 
In Section~\ref{section4} we  compute at one-loop  $\Delta_{m,q}$ 
 as functions of the SUSY breaking parameters and $\sqrt f$  and then present their 
numerical values in terms of the one-loop SM-like higgs mass.
For a most conservative case of low $\tan\beta$ and constrained MSSM 
boundary conditions for the soft terms, we find in ``non-linear'' MSSM 
an  ``acceptable''  $\Delta_m\approx 80$ ($\Delta_q\approx 120$) for $\sqrt f=2.8$ TeV and
$m_h\approx 126$ GeV.
This value of $\Delta$ can be reduced further for non-universal gaugino masses and 
 is well below that  in the constrained MSSM
(for any $\tan\beta$) where $\Delta_{m,q}\!\sim\! 800-1000$ \cite{KR}.
This reduction is done without enlarging the MSSM
spectrum (for an example with additional massive singlets see \cite{Ross:2011xv}).


\section{The Lagrangian in ``non-linear'' MSSM}\label{section2}

The  Lagrangian of the ``non-linear MSSM'' model can be written as
 \cite{nonlinearMSSM,Petersson:2011in,SK}
\medskip
\bea
\label{LL}
\cL=\cL_0+\cL_X+\cL_1+\cL_2
\eea
\medskip\noindent
$\cL_0$ is the usual MSSM SUSY Lagrangian which we write below to establish 
the notation:
\medskip
\bea\label{mssmsusy}
\cL_0 &=&\!\!\sum_{\Phi, H_{1,2}} \int d^4\theta \,\,
\Phi^\dagger\,e^{V_i}\,\Phi+
\bigg\{\int
d^2\theta\,\Big[\,\mu\,H_1\,H_2+
H_2\,Q\,U^c+Q\,D^c\,H_1+L\,E^c\,H_1\Big]+h.c.\bigg\}
\nonumber\\[-6pt]
&&
\hspace{1cm}+\sum_{i=1}^3\frac{1}{16\, g_i^2\,\kappa}
\int d^2\theta
\,\mbox{Tr}\,[\,W^\alpha\,W_\alpha]_i +h.c.,
\quad\quad \Phi:Q,D^c,U^c,E^c,L\, ,
\eea

\medskip\noindent
$\kappa$ is a constant canceling the trace factor and the
gauge coupling is $g_i$, $i=1,2,3$  for $U(1)_Y$, $SU(2)_L$, $SU(3)$ respectively.
Further, $\cL_X$ is the Lagrangian of the goldstino superfield $X=(\phi_X,\psi_X,F_X)$
that breaks SUSY spontaneously and
whose Weyl component is ``eaten'' by the gravitino (super-Higgs effect \cite{Grisaru:1982sr}).
$\cL_X$ can be written as \cite{SK}
\medskip
\bea
\cL_X=\int d^4\theta \, X^\dagger X +\Big\{\int d^2\theta\,f\,X+h.c.\Big\}
\qquad
{\rm with}
\qquad
X^2=0.
\eea

\medskip\noindent
The otherwise interaction-free $\cL_X$ when endowed with a  constraint 
$X^2=0$ \cite{SK,Rocek:1978nb,Lindstrom:1979kq} describes (onshell) the 
Akulov-Volkov Lagrangian of the goldstino  \cite{Volkov:1973ix},
see also \cite{VA1,VA2,VA3,VA4,VA5,VA6}, with non-linear SUSY. 
The constraint has a solution  $\phi_X=\psi_X\psi_X/(2F_X)$
that  projects (integrates) out the sgoldstino field which becomes 
massive and is appropriate for a low energy description
of SUSY breaking. %
Further, $\langle F_X\rangle\sim- f$  fixes the SUSY breaking scale ($\sqrt f$)
and the breaking is transmitted to the visible sector by
the couplings of  $X$ to the MSSM superfields, to generate
the usual SUSY breaking (effective) terms in $\cL_1+\cL_2$ (see below).
These couplings are commonly parametrized (onshell)  in terms of the spurion
field $S=m_{soft}\theta\theta$ where $m_{soft}$ is a generic notation for the soft 
masses (later denoted $m_{1,2,3}$, $m_{\lambda_i}$); however,  this parametrisation 
obscures the  dynamics of $X$  (offshell effects)  relevant below that generates 
additional Feynman diagrams mediated by  $F_X$ (Figure~\ref{fig1}).
 Such effects  are not seen in the leading order (in $1/f$) in the  spurion formalism. 
The offshell couplings are easily recovered by  the formal replacement \cite{SK}
\medskip
\bea
S\!\ra\! \frac{m_{soft}}{f} X\,\,
\eea

\medskip\noindent
In this way one obtains the SUSY breaking couplings that are indeed 
identical to those obtained by the equivalence theorem \cite{cddfg} from 
a theory with the corresponding explicit soft breaking terms and in which
the Goldstino fermion couples to the derivative of the supercurrent of 
the initial theory. These couplings are generated by  the D-terms below
\medskip
\bea
\cL_{1}\!\! &=&\!\!\sum_{i=1,2} c_i
\int d^4\theta \,\,X^\dagger X\,\,
H_i^\dagger\,e^{V_i}\,H_i
+
\sum_{\Phi} c_\Phi\int d^4\theta \,\,
X^\dagger X\,\Phi^\dagger
e^V\,\Phi.
\eea
and by the F-terms:
\bea
\cL_{2}
\!\!&=&
\sum_{i=1}^3
\frac{1}{16\, g^2_i\,\kappa}
\frac{2\,m_{\lambda_i}}{f}
\int d^2\theta
\,X\,\mbox{Tr}\,[\,W^\alpha\,W_\alpha]_i 
+
c_3\int d^2\theta \,X\,H_1\,H_2
\nonumber\\[2pt]
&+&
\frac{A_u}{f}\int d^2\theta\,X\,H_2\,Q\,U^c+
\frac{A_d}{f}\,\int d^2\theta\,X\,Q\,D^c\,H_1+
\frac{A_e}{f}\,\int d^2\theta\,X\,L\,E^c\,H_1+h.c.
\label{as}
\eea
with
\bea
c_{j}=-\frac{m_j^2}{f^2},\,\,j=1,2;
\qquad 
c_3=-\frac{m_3^2}{f},
\qquad
c_\Phi=-\frac{m_\Phi^2}{f^2},\qquad \Phi: Q, U^c, D^c, L, E^c,
\eea

\medskip\noindent
In the UV one can eventually take $m_\Phi=m_0=m_1=m_2$, $m_{\lambda_i}=m_{12}$ ($i=1,2,3$)
for all gaugino masses, $m_3^2=B_0\,m_0\,\mu_0$,  ($\mu\equiv \mu_0$ in the UV)
and these define the ``constrained'' version of the ``non-linear'' MSSM,
discussed later.
For simplicity, Yukawa matrices are not displayed; to recover them
just replace above any pair of
fields $\phi_Q \phi_U\ra \phi_Q\gamma_u \phi_U$,
$\phi_Q\phi_D\ra \phi_Q\gamma_d \phi_D$,
$\phi_L\,\phi_E\ra \phi_L\gamma_e \phi_E$;
similar for the fermions and auxiliary fields,
 with $\gamma_{u,d,e}$ $3\times 3$ matrices.

The total Lagrangian $\cL$ defines the model discussed in  detail
in \cite{nonlinearMSSM}.
The only difference from the ordinary MSSM is in the  supersymmetry breaking 
sector.  In the calculation of the onshell Lagrangian
we restrict the calculations to up to and including $1/f^2$ terms.
This requires solving for $F_\phi$ of matter fields
up to and including  $1/f^2$ terms and for
$F_X$ up to and including $1/f^3$ terms (due to its leading
contribution which is -$f$).
In this situation, in the final Lagrangian no kinetic mixing is present
at the order used\footnote{We stress that at energy scales below $m_{soft}$,
similar constraints to that used for $X$ ($X^2=0$)  
can be applied to the MSSM superfields themselves and
 correspond to integrating out the massive superpartners \cite{SK}.}.


\section{The Higgs masses at one-loop in ``non-linear'' MSSM}\label{section3}

From the Lagrangian  $\cL$ one obtains the Higgs scalar potential of the 
model\footnote{
In the standard notation for a two-higgs doublet model
$V=\tilde m_1^2\,\vert h_1\,\vert^2
+\tilde m_2^2\,\vert h_2\,\vert^2
-(m_3^2\,h_1 \cdot h_2+h.c.)+
\frac{1}{2}\,\lambda_1 \,\vert h_1\,\vert^4
+\frac{1}{2}\,\lambda_2 \,\vert h_2\,\vert^4
+\lambda_3 \,\vert h_1\,\vert^2\,\,\vert h_2\,\vert^2\,
+\lambda_4\,\vert\,h_1\cdot h_2\,\vert^2
+ \Big[
\frac{1}{2}\,\lambda_5\,(h_1\cdot  h_2)^2+\lambda_6\,\vert\,h_1\,\vert^2\, 
(h_1 \cdot h_2)+
\lambda_7\,\vert\,h_2\,\vert^2\,(h_1 \cdot h_2)+h.c.\Big]$
where $\tilde m_1^2=m_1^2+\vert \mu\vert^2$, $\tilde m_2^2=m_2^2+\vert \mu\vert^2$.
$ \lambda_1/2=g^2/8+m_1^4/f^2$,
$\lambda_2/2=g^2\,(1+\delta)/8+m_2^4/f^2$,
$\lambda_3=(g_2^2-g_1^{2})/4+2\, m_1^2 m_2^2/f^2$,
$\lambda_4= -g_2^2/2+m_3^4/f^2$,
$\lambda_5=0$, 
$\lambda_6=-m_1^2 m_3^2/f^2$,
$\lambda_7=-m_2^2 m_3^2/f^2$, $g^2=g_1^2+g_2^2$.}
\bea
\label{potential0}
V&=&
\big(\vert \mu\vert^2+m_1^2\big)\,\,
\vert h_1\vert^2+
\big(\vert \mu\vert^2 +m_2^2\big)
\vert h_2\vert^2
-\big(m_3^2\,h_1.h_2+h.c.\big)
\\[4pt]
&+&\!\!\!
\frac{1}{f^2}\,\Big\vert m_1^2\,\vert h_1\vert^2+m_2^2\,\vert h_2\vert^2
- m_3^2\,h_1.h_2\Big\vert^2
+\frac{g_1^2+g_2^2}{8}\,\Big[\vert h_1\vert^2-\vert h_2\vert^2\Big]^2
+\frac{g_2^2}{2}\,\vert h_1^\dagger\,h_2\vert^2
\nonumber\\[4pt]
&+&
\frac{g_1^2+g_2^2)}{8}\,\delta \,\vert h_2\vert^4
+\cO(1/f^3)\nonumber
\eea
with $h_1.h_2=h_1^0 h_2^0-h_1^- h_2^+$, 
$\vert h_1\vert^2=\vert h_1^0\vert^2+\vert h_1^-\vert^2$,
$\vert h_2\vert^2=\vert h_2^0\vert^2+\vert h_2^+\vert^2$.

\smallskip
\begin{figure}[t!]
\centering\def\baselinestretch{1.}
  \includegraphics[height=0.14\textheight,width=0.85\textwidth]{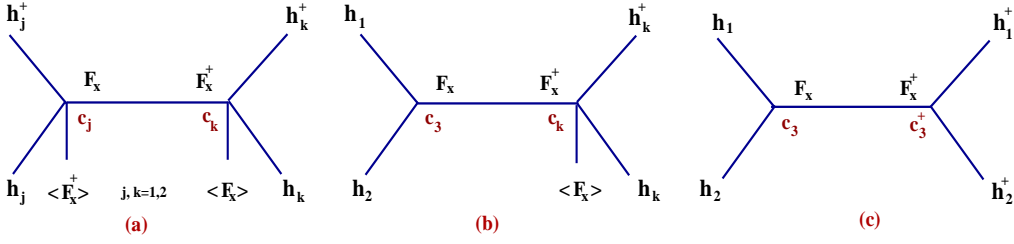}
\caption{{\protect\small
The diagrams  that generate the new quartic effective
 higgs couplings in V, eq.(\ref{potential0}).
The coefficients $c_{1,2,3}$ are generated by $\cL_1$, $\cL_2$. 
$F_X$ is the auxiliary component of $X$  that breaks SUSY. The left
 (right) diagrams are generated by D (F) terms in the action,
while the middle one is a mixture of both. 
These  interactions are generic and important in low-scale SUSY breaking models
 \cite{Casas:2003jx,Brignole:2003cm}  (in   MSSM they are strongly suppressed 
since $\langle F_X\rangle$  is large).}}
\label{fig1}
\end{figure}

What is interesting in the above higgs potential is the
presence of the first term in the second line of $V$, 
 absent in MSSM, that is generated by the diagrams in Figure~\ref{fig1}.
 Therefore, quartic higgs terms are generated by the dynamics of the 
goldstino superfield and are not captured 
 by the usual spurion formalism in the MSSM. 
The impact of these terms for phenomenology is important and analyzed below, for when 
$\sqrt f\sim $ few TeV, see  \cite{Casas:2003jx,Brignole:2003cm} for a related study.  
When $\sqrt f$ is very large which is the case 
of MSSM-like  models, these terms are negligible and thus
not included by the spurion formalism.
The ignored higher order terms $\cO(1/f^3)$  involve non-renormalizable  
$h_{1,2}^6$ interactions in $V$ and are not considered 
here\footnote{Effective operators in the Higgs sector in the SUSY 
context were discussed in the past \cite{Brignole:2003cm, Cassel:2009ps, Carena:2009gx}.}.
Finally, the radiatively corrected $m_{1,2,3}$ and $\mu$ in $V$ depend on the 
scale (hereafter denoted $t$) while
the term $\delta \vert h_2\vert^4$ is generated at one-loop by top-stop Yukawa couplings.
We thus neglect other Yukawa couplings and our one-loop analysis is valid for low $\tan\beta$;
including two-loop  leading log effects $\delta$ is
\bea
\delta=\frac{3 \,h_t^4}{g^2\pi^2}
\Big\{\ln\frac{M_{\tilde t}}{m_t}+ \frac{X_t}{4}
+\frac{1}{32\pi^2}\,(3\,h_t^2-16\,g_3^2)\,\Big(X_t+2\ln \frac{M_{\tilde t}}{m_t}\Big)
\ln \frac{M_{\tilde t}}{m_t}\Big\}
\eea
where
\bea
X_t\equiv \frac{2 (A_t \,m_0-\mu\cot\beta)^2}{M_{\tilde t}^2}
\Big(1-\frac{(A_t \,m_0-\mu\cot\beta)^2}{12 M_{\tilde t}^2}\Big),
\eea

\medskip\noindent
 $M_{\tilde t}^2=m_{\tilde t_1}\,m_{\tilde t_2}$ and $g_3$ is the QCD coupling and $A_t$ is the
dimensionless trilinear top coupling\footnote{
More exactly  $A_t=A_u/m_0$ 
with $A_u$ as in eq.(\ref{as}).}.

The minimum conditions of the potential can be written
\medskip
\bea\label{min}
-v^2=\frac{m^2}{\lambda},\qquad 
2\,\lambda\,\frac{\partial
  m^2}{\partial\beta}-m^2\,\frac{\partial\lambda}{\partial \beta}
=0
\eea
with the notation\footnote{Also
$\lambda\equiv  (\lambda_1/2) c_\beta^4+ (\lambda_2/2)\,s_\beta^4
+(\lambda_3+\lambda_4+\lambda_5)\,s_\beta^2\,c_\beta^2
+2\lambda_6\,c_\beta^3\,s_\beta+ 2\lambda_7\,c_\beta\,s_\beta^3
$ where we used the notation of  footnote 3 and
$s_\beta=\sin\beta$, $c_\beta=\cos\beta$, $u\equiv
\tan\beta=v_2/v_1$, $h_i=1/\sqrt{2}\,(v_i+\tilde h_i)$, 
$m_Z^2=(g_1^2+g_2^2)\,v^2/4$.}:
\medskip
\bea 
m^2&\equiv &
(m_1^2+\mu^2)\,\cos^2\beta+(m_2^2+\mu^2)\,\sin^2 \beta-m_3^2\,\sin 2\beta 
\nonumber\\[6pt]
\lambda
&\equiv& \frac{g_1^2+g_2^2}{8}
\Big[\cos^2 2\beta+\delta \sin^4 \beta\Big]
+\frac{1}{f^2}\,\Big\vert
m_1^2\cos^2\beta+m_2^2 \sin^2 \beta -(1/2) \,m_3^2\,\sin  2\beta
\Big\vert^2
\eea

\medskip\noindent
The correction to the effective quartic higgs coupling
$\lambda$, due to the soft terms ($m_{1,2,3}$) 
has implications for the higgs mass and EW fine tuning. 
This positive correction could
alleviate the relation between $v^2$ and $m^2$: indeed, with $m\sim \cO$(1 TeV) and
$\lambda\sim \cO(1)$, $v$ can only be of order $\cO(1\,$TeV$)$ as well. This 
brings a tension between the EW scale and soft terms ($\sim m$) which cannot easily
be separated from each other; this tension is encoded by 
 the EW fine-tuning measures, discussed in Section~\ref{section4}.
Increasing $\lambda$  can alleviate this tension, with impact 
on the EW fine tuning.
Such correction to $\lambda$  also arises 
in models with high scale breaking in the 
hidden sector, so it is present even in usual MSSM  but is
extremely small in that case since then $\sqrt f\sim 10^{10}$ GeV. 
Here we consider $\sqrt f\sim$~few TeV, which is safely above the current lower 
 bound of $\approx 700$ GeV \cite{nonlinearMSSM,Brignole:2003cm,lowerbound,Antoniadis:2004se}.

The two minimum conditions of the scalar potential lead to:
\medskip
\bea\label{ttt}
 m_1^2-m_2^2&=&
\cot 2\beta\,\bigg[\,-m_3^2+\frac{f^2}{v^2}
\frac{(-1+  \sqrt w_0)\,[
\,m_3^2 + m_Z^2\,\sin 2\beta\,\big(1-(\delta \sin^2\beta)/(2\cos 2\beta)\big) \, ]}
{2\,\mu^2+m_Z^2 (\cos^2 2\beta+ \delta \sin^4\beta)-m_3^2\,\sin 2\beta}\,\bigg]
\nonumber\\[10pt]
m_1^2+m_2^2&=&
\frac{1}{\sin 2\beta}\,\bigg[
m_3^2 +\frac{f^2}{v^2}\frac{(-1+\sqrt w_0)\,[
-m_3^2 +\big(2\mu^2 +(\delta/2) m_Z^2 \sin^2\beta\big)\,
 \sin 2\beta ]
}{
2\mu^2+m_Z^2\, ( \cos^2 2\beta+\delta \sin^4\beta)-m_3^2\sin 2\beta}\,\bigg]
\eea
%
where:
\bea
w_0\equiv 1-\frac{v^2}{f^2}\,\big(4\mu^2+2\,m_Z^2\,(\cos^2 2\beta+\delta\,\sin^4\beta)
-2\,m_3^2\,\sin 2\beta\big)
\eea
There is a second solution for $m^2_{1,2}$  at the minimum (with minus
in front of $\sqrt w_0$)
which however is not a perturbation of the MSSM solution and is not considered
below (since it brings a shift proportional to $f$ of the soft masses, 
which invalidates the expansion in $m_{1,2}^2/f$).

The mass of the  pseudoscalar higgs is, 
including  a one-loop correction (due to $\delta$):
\medskip
\bea
m_A^2=\frac{2 m_3^2}{\sin 2\beta}\,
\Big\{
\frac{3+\sqrt w_0}{4}-\frac{m_3^2 \,v^2}{4 f^2}\,\sin 2\beta\Big\}
\eea

\medskip\noindent
which can be expanded to $\cO(1/f^3)$ using the expression of $w_0$. For large $f$
one recovers its MSSM expression at one-loop.
Further, we computed the masses $m_{h,H}$  including the one-loop correction (due to $\delta$) to find:
\bea\label{mhH}
m_{h,H}^2& =&
\frac{1}{2}\Big[ m_A^2+m_Z^2\mp \sqrt w 
+
\delta\,m_Z^2\,\sin^2\beta\Big]+\Delta m^2_{h,H}
\eea

\medskip\noindent
with upper (lower) sign corresponding to $m_h$ ($m_H$) and the correction $\Delta m_{h,H}^2=
\cO(1/f^2)$ is:
\bea\label{deltamhH}
\Delta m^2_{h,H}\!\!\!\!&=&\!\!\!
\frac{v^2}{64\,f^2}
\Big\{
8 \, \big[\, 8\mu^4
-2 \,m_A^2\,\mu^2+4\mu^2\,m_Z^2+m_Z^4
+
(2\,m_A^2\,\mu^2+4\,\mu^2\,m_Z^2+m_Z^4)\cos 4\beta\,\big]
\nonumber\\
&-&
16\,\delta\, m_Z^2\,\big[ m_A^2-4\mu^2+(m_A^2+2m_Z^2)\cos 2\beta 
\,\big]\sin^4\beta
+
16\,\delta^2\,m_Z^4\sin^6\beta
\nonumber\\  
&\pm &\!\! (1/\sqrt w)
\Big[
3\, m_A^6-m_A^4 (16\mu^2+m_Z^2)+ 4 m_A^2 (16\mu^4+4\mu^2 \,m_Z^2+m_Z^4)
- 8 m_Z^4 (4 \mu^2 +m_Z^2)
\nonumber\\  [-1pt]
&-&\!\! 4 \big[ 
m_A^6+m_A^4(m_Z^2-4\mu^2)-2 m_A^2 m_Z^2(6\mu^2+m_Z^2)
+
 2 m_Z^2 (8\mu^4\!+\! 4\mu^2 m_Z^2 \!+ \! m_Z^4)\big]\cos 4\beta
\nonumber\\ 
&+&
m_A^2\, (m_A^2+m_Z^2)(m_A^2+4m_Z^2)\cos 8\beta
+4\,\delta\, m_Z^2
\,\big[-m_A^4 - 2 m_Z^4 +m_A^2 (8\mu^2+m_Z^2)
\nonumber\\ 
&+&
\big( (m_A^2-4\mu^2)^2 -3 (m_A^2-8\mu^2) m_Z^2+7m_Z^4\big)\cos 2\beta
+
\big(m_A^4+(3m_A^2-8\mu^2)m_Z^2
\nonumber\\ 
& -& 2 m_Z^4\big)\cos 4\beta -
(m_A^4+m_A^2 m_Z^2 -m_Z^4)\cos 6\beta\big]\sin^2\beta
+
16\,\delta^2\,m_Z^4\, (m_A^2-4\mu^2
\nonumber\\ 
&+& 
3 \,m_Z^2 \cos 2\beta)\sin^6\beta-16\,\delta^3\,m_Z^6\sin^8\beta
\Big]\Big\}+\cO(1/f^3).
\eea
with
\bea
w\equiv (m_A^2+m_Z^2)^2-4\,m_A^2\,m_Z^2\,\cos^2 2\beta
+2\,\delta\,(m_A^2-m_Z^2)\,m_Z^2\,\cos(2\beta)\,\sin^2\beta
+\delta^2\,m_Z^4\sin^4\beta
\eea

\medskip\noindent
It is illustrative to take the limit of large $\tan\beta$ on $m_{h,H}^2$
with $m_A$ fixed. One finds
\bea\label{lim}
m_{h}^2&=&\big[(1+\delta) m_Z^2
+ \frac{v^2}{2 f^2} \big(2 \mu^2+ (1+\delta)\, m_Z^2\big)^2+\cO(\cot^2\beta)\big]
 +\cO(1/f^3),
\nonumber\\
m_H^2&=& \big[m_A^2+\cO(\cot^2\beta)\big]+\cO(1/f^3).
\eea

\medskip\noindent
where we ignored the $\tan\beta$ dependence of $\delta$. Due to the 
$\cO(\cot^2\beta)$ suppression,  eq.(\ref{lim}) is valid even at smaller 
$\tan\beta\sim 10$. In this limit a significant increase  of $m_h$ to 
$120$ or even $126$ GeV is easily achieved,  driven  by classical effects 
alone with  $\mu$  near TeV (and eventually small quantum corrections, $\delta\sim 0.5$). 
Such  increase due to $\mu$ is thus of SUSY origin, even though the quartic Higgs 
couplings ($\cO(1/f^2)$)  giving this effect involved the soft masses $m_{1,2,3}$.  
These  combined to give, at the EW minimum,  the $\mu$-dependent increase in
eq.(\ref{lim}). For large $f$ one recovers the MSSM value of 
$m_{h,H}$, at one loop. Eqs.(\ref{mhH}), (\ref{deltamhH}) are used in  
Section~\ref{section4} to analyze the EW fine-tuning as a function of $m_h$.


\section{The electroweak scale fine tuning}\label{section4}

\subsection{General results}\label{gg}

To compute the EW fine tuning we use two definitions for it 
already shown in Introduction:
\bea\label{ss}
\Delta_m=\max\big\vert \Delta_{\gamma^2}\big\vert,
\quad
\Delta_q=\Big\{\sum_{\gamma} \Delta_{\gamma^2}^2\Big\}^{1/2}, 
\,\,
{\rm with}
\,\,
\Delta_{\gamma^2}\!\equiv\!\frac{\partial \ln v^2}{\partial\ln\gamma^2}, 
\eea
\noindent
where $\gamma= m_0, m_{12}, A_t, B_0, \mu_0$ for the constrained ``non-linear'' MSSM.
In the following we evaluate $\Delta_m$, $\Delta_q$ at the one-loop level in our model.
Using eqs.(\ref{min})
 that give $m^2=m^2(\gamma, \beta)$ and $\lambda=\lambda(\gamma,\beta)$
one has a general result for $\Delta_{\gamma^2}$ which takes into account that
$\tan\beta$ depends on $\gamma$ via the second min condition in eq.(\ref{min}). 
The result is
\cite{Casas:2003jx} 
\medskip 
\begin{equation} 
\Delta _{\gamma^2}=-\frac{\gamma}{2\,z}\,\bigg[\bigg(2\frac{\partial ^{2}m^{2}}{\partial 
\beta ^{2}}+v^{2}\frac{\partial ^{2}\lambda }{\partial \beta ^{2}}\bigg)
\bigg(\frac{\partial \lambda }{\partial \gamma}+\frac{1}{v^{2}}\frac{\partial 
m^{2}}{\partial \gamma}\bigg)+\frac{\partial m^{2}}{\partial \beta }\frac{
\partial ^{2}\lambda }{\partial \beta \partial \gamma}-\frac{\partial \lambda }{
\partial \beta }\frac{\partial ^{2}m^{2}}{\partial \beta \partial \gamma}\bigg]. 
\label{delta2} 
\end{equation} 
where
\bea\label{zz}
z\equiv \lambda \,\bigg(2\frac{\partial ^{2}m^{2}}
{\partial \beta ^{2}}+v^{2}\frac{ 
\partial ^{2}\lambda }{\partial \beta ^{2}}\bigg)-\frac{v^{2}}{2}\,\bigg(
\frac{\partial \lambda }{\partial \beta }\bigg)^{2}. 
\eea

\medskip\noindent
Using these expressions, one obtains  $\Delta_m$ and $\Delta_q$.

Let us first consider the limit of large $\tan\beta$, so the first relation in eq.(\ref{min})
becomes
\medskip
\bea
v^2=-\frac{2 \,(m_2^2+\mu^2)}{(1+\delta)\,(g_1^2+g_2^2)/4+2\,m_2^4/f^2}+\cO(\cot\beta)
\eea
which gives
\bea\label{kk}
\Delta_{\gamma^2}=-\frac{\partial (m_2^2+\mu^2)}{\partial\ln\gamma}
\,\frac{(1+2 \,v^2\,m_2^2/f^2)^s}{(1+\delta)\,m_Z^2+2\,v^2\,m_2^4/f^2}
+\cO(\cot\beta),
\eea

\medskip\noindent
where $s=1\,{\rm if}\,\gamma\not=\mu_0;\,\, s=0\,{\rm if}\,\gamma=\mu_0$,
and $\mu$, $m_2$ are functions of 
the scale\footnote{as we shall detail shortly  for the case of the constrained MSSM.}.
If also $f$ is large, one recovers the MSSM corresponding expression 
(ignoring a $\tan\beta$ dependence of $\delta$):
\medskip
\bea
\Delta^0_{\gamma^2}=-\frac{\partial (m_2^2+\mu^2)}{\partial\ln\gamma}
\,\frac{1}{(1+\delta)\,m_Z^2}
+\cO(\cot\beta).
\eea

\medskip\noindent
which is interesting on its own.
For the EW symmetry breaking to exist one must have $m_2^2+\mu^2<0$ 
and  therefore $\Delta_{\gamma^2}$ of the ``nonlinear MSSM'' 
is smaller than in the  MSSM with similar UV boundary conditions
for parameters
$\gamma$.
Indeed, in this case
the ratio $r$ of $\Delta_{\gamma^2}$ to that in a MSSM-like model denoted
$\Delta_{\gamma^2}^0$:
\medskip
\bea
r=\frac{\Delta_{\gamma^2}}{\Delta_{\gamma^2}^0}=
\frac{(1+2 \,v^2 \,m_2^2/f^2)^s\,(1+\delta)\,m_Z^2}{(1+\delta)\,m_Z^2+2\,v^2\,m_2^4/f^2}
+\cO(\cot\beta), 
\eea

\medskip\noindent
is smaller than unity: $r\approx 1/2$ if $\delta\approx 0.8$, 
$\vert m_2^2\vert/f\approx 0.35$
and $r\approx 1/3$ if $\delta\approx 0.8$, $\vert m^2_2\vert/f\approx 0.5$ 
with $\sqrt f$ above the TeV scale
(recall $\vert m_2^2\vert /f<1$ for convergence and $\delta\sim 0.5-1$).
So for a large $\tan\beta$  the EW fine tuning associated to each 
UV parameter is smaller  relative to
 the MSSM and the same can then be said about overall $\Delta_m$ and $\Delta_q$. 
This reduction is actually more significant,  since for the same point in the
parameter space the higgs mass is larger in the ``nonlinear'' MSSM than in the MSSM alone,
already at the tree level.
Indeed, we saw in eq.(\ref{lim}) that even in the absence of loop corrections
one can easily achieve $m_h\approx 120$ GeV, without the additional,
significant fine-tuning ``cost'', present for $m_h>115$ GeV in the MSSM.
This ``cost'' is  $\Delta\sim \exp(\delta m_h/{\rm GeV})$ 
due to loop corrections needed to increase $m_h$ by $\delta m_g$ in MSSM models\footnote{
For this exponential dependence on $m_h$ see figures 1 and 6 in the first reference
in \cite{Cassel:2010px}.}; 
for the same $m_h$ the reduction is then expected to be by a factor 
$\Delta\sim \exp (120-115)\sim 150$  relative to the constrained MSSM case.
Then our $\Delta_{m,q}$ can be smaller by this factor 
and  $r$ is also much smaller than unity when evaluated for the same $m_h$.
Finally, fixing $m_h$ to its measured value is a very strong 
constraint on the parameter space, which once satisfied, allows  other EW constraints 
to be automatically respected \cite{KR}, so this conclusion is unlikely to be affected
by them.

Let us mention that in MSSM-like models the EW fine-tuning $\Delta$
is usually reduced as one increases $\tan\beta$ for a fixed $m_h$
(all the other parameters allowed to vary) \cite{Cassel:2010px}. 
This is because  at large $\tan\beta$ additional Yukawa couplings effects (down sector) are enhanced
and help the radiative EW symmetry breaking (thus reducing $\Delta$), 
while at small $\tan\beta$ this effect is suppressed \cite{KR}. The situation is similar in the
above ``nonlinear'' MSSM model\footnote{As
 we show shortly for the conservative case  of the constrained ``non-linear'' MSSM, at
 small $\tan\beta$, fine tuning is already acceptable, thus 
at larger $\tan\beta$ $\Delta$ is expected to be similar or further reduced.}.

\subsection{The constrained ``non-linear'' MSSM}

The reduction of the EW fine tuning in our model 
can be illustrated further  by comparing it with that in the  constrained MSSM (CMSSM)
with universal UV scalar mass $m_0$ and gaugino mass $m_{12}$ and including
only the top/stop Yukawa coupling correction. In that case one has
\medskip
\bea\label{ms}
m_1^2(t)
& =& m_0^2
 +m_{12}^2\,\sigma_1(t),\qquad
\mu^2(t)=\mu_0^2\,\sigma_8^2(t)
\nonumber\\
m_2^2(t)
&=& 
m_{12}^2\,\sigma_4(t) +A_t\,m_0\,m_{12}\,\sigma_5(t)
+m_0^2\,\sigma_7(t) - m_0^2\,A_t^2\,\sigma_6(t)
\nonumber\\
m_3^2(t) &=& \mu_0\,m_{12}\,\sigma_2(t)+ B_0\,m_0\,\mu_0\,\sigma_8(t)+
\mu_0\,m_0\,A_t\,\sigma_3(t)
\eea

\medskip\noindent
where we made explicit the dependence of soft masses  $m_{1,2,3}$ and  $\mu$ and 
of the coefficients
$\sigma_i$  on the momentum scale $t=\ln \Lambda_{UV}^2/q^2$ induced by radiative corrections; 
$\sigma_i$ also depend   on $\tan\beta$ and so do the soft masses. The high 
scale boundary conditions are chosen such as $\sigma_{1,2,3,4,5,6}(0)\!=\!0$,
$\sigma_{7,8}(0)\!=\!1$ when quantum corrections are turned off.
 For $q^2\!=\!m_Z^2$ the values of $\sigma_i$  are given
in the Appendix. These expressions are used in our numerical analysis below.

 \subsubsection{The large $\tan\beta$ case}

This regime was already discussed in the general case in Section~\ref{gg}.
A numerical analysis of this case involves additional Yukawa couplings of the ``down'' sector
not included in our $V$ and  is beyond the goal of this paper. 
However, we can still provide  further insight  for the constrained ``nonlinear MSSM''.
From eq.(\ref{kk}), one has
\bea
\!\!\!\Delta_{\mu_0^2}&=&-\frac{2\mu_0^2\,\sigma_8^2}{(1+\delta)\,m_Z^2
+2\,v^2 m_2^4/f^2}+\cO(\cot^2\beta)
\nonumber\\
\Delta_{m_0^2}&=&-\frac{m_0\,(1+2 v^2 m_2^2/f^2)}{(1+\delta)\,m_Z^2+2\,v^2\,m_2^4/f^2}
(A_t\sigma_5-2 A_t^2\,m_0\sigma_6+2 m_0\sigma_7)+\cO(\cot\beta)
\nonumber\\
\Delta_{m_{12}^2}&=&-\frac{m_{12}\,(1+2 v^2 m_2^2/f^2)}{(1+\delta)\,m_Z^2+2\,v^2\,m_2^4/f^2}
(2 m_{12}\sigma_4 +A_t\,m_0 \sigma_5)+\cO(\cot\beta)
\nonumber\\
\Delta_{A_t^2}&=&-\frac{A_t\,(1+2 v^2 m_2^2/f^2)}{(1+\delta)\,m_Z^2+2\,v^2\,m_2^4/f^2}
(m_{12}\sigma_5-2 m_0\,A_t\sigma_6)\,m_0+\cO(\cot\beta), \,\,
\nonumber\\
\Delta_{B_0^2}&=& \cO(\cot\beta)~~~
\label{rr}
\eea
 $m_2^2$ is given in eqs.(\ref{ms}) and
with $m_2^2<0$, the absolute values of above  $\Delta$'s and then of $\Delta_{m,q}$
are smaller than those in the 
limit $f\ra \infty$ when one recovers the  constrained MSSM model (at large
$\tan\beta$).
So fine tuning is reduced as already argued in the general discussion.

Turning off the quantum corrections to 
 soft masses and $\mu$ ($\sigma_{1,2,..,6}\!=\!0$, $\sigma_{7,8}\!=\!1$) and 
quartic coupling ($\delta\!=\!0$), for large $f$, 
the above relations simplify to give for constrained MSSM
\bea\label{erer}
\vert \Delta_{\gamma^2}\vert=\frac{2 \gamma^2}{m_Z^2}+\cO(\cot\beta), \qquad\gamma=m_0, \mu_0
\eea
with remaining expressions being $\cO(\cot\beta)$.
This also shows that in the constrained MSSM, the dominant contributions to fine tuning
(at classical level) are due to $m_0$ and $\mu_0$. In general $\Delta_{m_0^2}$ 
 is related to QCD effects that increase fine tuning and dominates 
 for $m_h\!>\!115$ GeV (fig.2 in first reference in \cite{Cassel:2010px}). For TeV-valued
$m_0=\mu_0=2$ TeV ($\delta=0$) one then has $\Delta_{q}=683$ which gives a good
estimate of the value of fine tuning in constrained MSSM\footnote{For $m_h\approx 126$ GeV, 
in constrained MSSM $\Delta_{m,q}\sim 800-1000$ \cite{KR}.}.
Eq.(\ref{erer}) has close similarities to other fine-tuning measures defined in the literature
such as $\Delta_{EW}$ of \cite{HB}.

\subsubsection{The small $\tan\beta$ case}

From eqs.(\ref{ss}), (\ref{delta2}), (\ref{zz}) we find the following analytical 
results for $\Delta_{\gamma^2}$ at one loop level:
\bea\label{dmu}
\Delta_{\mu_0^2}\!&\!=&\!-\frac{4}{D\,v^2} \Big\{-2 f^2 y_1 \sin2\beta 
\Big[ (4+\delta) f^2 m_Z^2 + 2 v^2 (y_1^2+ y_2^2) 
  - 2 (\delta f^2 m_Z^2 +v^2 \,y_2 \,y_3) \cos 2\beta
\nonumber\\
&+&\!\!
\big[ (4+\delta) f^2 m_Z^2\! +2 v^2 y_1^2 \big]\cos 4\beta\! -2 v^2 y_1 y_2 \sin 4\beta
\Big]\!+\!
\Big[
\big[ f^2 (m_Z^2\delta\! +4 y_2)\!+\! 2 v^2 y_2 y_3\big] \cos 2\beta
\nonumber\\
&-& \!\!
\big[ (4+\delta) f^2 m_Z^2 +2 v^2  (-y_1^2+ y_2^2)\big]\cos 4\beta
+
2 y_1 (4 f^2+v^2 y_3 -4 v^2 y_2 \cos 2\beta) \sin 2 \beta
\Big]
\nonumber\\
&\times&\!\!\!\!
\Big[
8 f^2 \mu_0^2 \si_8^2 +v^2 y_1^2 + y_1  \big[- 4 f^2 \sin 2\beta 
+
v^2 (-y_1 \cos 4\beta -2 y_3\sin 2\beta +y_2 \sin 4\beta)\big]\Big]\Big\}
\eea
%
\bea
\Delta_{m_0^2}\!\!\!&=&
\!\! - \frac{4 f^2 m_0}{D}
\Big\{ 4 \Big[ 
\big[ f^2 (m_Z^2\delta +4 y_2)\! +\!2 v^2 y_2 y_3\big]\cos 2\beta
\!-
\big[ (4\!+\!\delta ) f^2 m_Z^2\! +\! 2 v^2 (y_2^2\!-y_1^2)\big] \cos 4\beta
\nonumber\\
&+&\!\!\!\!\!\!
2 y_1 \big[4 f^2\!\! +\! v^2 (y_3\! - \! 4  y_2 \cos 2\beta) \big]\sin 2\beta 
\Big]
\Big[
v^{-2} \big[ 2 m_0 \cos^2\!\beta\!+\! y_4 \sin^2\!\beta 
\! - \! \mu_0 (A_t \sigma_3\! +\!\! B_0\sigma_8) \sin 2\beta\big]
\nonumber\\
&+&\!\!\!\!\!
(1/f^2) \big[ 2 m_0 \cos^2\beta\! -\!\mu_0 (A_t \sigma_3\!
 +B_0\sigma_8)\cos\beta\sin\beta\! +y_4 \sin^2\beta\big]
(y_3\!-\! y_2 \cos 2\beta\!-\! y_1 \sin 2\beta)
\Big]
\nonumber\\
&+&
8\, (- 2 y_1 \cos 2\beta + y_2 \sin 2\beta)
\Big[
(1/2)  \big[\mu_0 (A_t \sigma_3 + B_0\sigma_8 ) \cos 2\beta 
+ (2 m_0 -y_4) \sin 2\beta\big]
\nonumber\\
&\times &(y_2 \cos 2\beta-y_3+y_1\sin 2\beta)
\! -\!
\big[ 2 m_0 \cos^2\beta - \mu_0 (A_t \sigma_3\! +\! B_0 \sigma_8 )(1/2) \sin2\beta
\!+\! y_4 \sin^2\beta\big]
\nonumber\\
&\times & (y_1 \cos 2\beta -y_2 \sin2 \beta)
\Big]
+
(1/v^2) \big[ 2\mu_0 (A_t \sigma_3
 + B_0 \sigma_8)  \cos 2\beta+ (2m_0-y_4) \sin 2\beta\big]
\nonumber\\
&\times &
\big[
-2 f^2 m_Z^2 (-\delta +(4+\delta)\cos 2\beta)\sin2\beta
+4 v^2 (-y_3 +y_2 cos 2\beta + y_1\sin 2\beta )(y_1\cos 2\beta \nonumber\\
&-&\!\! y_2 \sin 2\beta)
\big]
\Big\}
\eea
and
\bea
\Delta_{m_{12}^2}&=&
\frac{-4 f^2 m_{12}}{D} \Big\{
4 \Big[ \big[f^2 (m_Z^2\delta +4 y_2)+ 2 v^2 y_2 y_3\big]\cos 2\beta
-
\big[ (4+\delta) f^2 m_Z^2 + 2 v^2 (y_2^2-y_1^2) \big]
\nonumber\\
&\times&\!\! \cos 4\beta
\!+\!
2 y_1 ( 4 f^2 +v^2 y_3 -4 v^2 y_2 \cos 2\beta ) \sin 2\beta 
\Big]
\Big[
\frac{1}{v^2} 
\big[ 2 m_{12} \sigma_1 \cos^2\beta - \mu_0 \sigma_2 \sin2\beta
\nonumber\\
&+&
 (2 m_{12} \sigma_4 + A_t m_0 \sigma_5 )\sin^2\beta \big]
+
(1/f^2) \big[ 2 m_{12} \sigma_1 \cos^2\beta - (1/2)\,\mu_0 \sigma_2 \sin2\beta
\nonumber\\
&+& (2 m_{12} \sigma_4 + A_t m_0 \sigma_5 )\sin^2\beta \big]
(y_3-y_2 \cos2\beta - y_1 \sin 2\beta)
\Big]
+
8 \,(y_2\sin 2\beta -2 y_1 \cos 2\beta) 
\nonumber\\
&\times&\!\!\!
\Big[
(1/2) \big[\mu_0 \sigma_2 \cos 2\beta\! +\! \big(2 m_{12} (\sigma_1-\sigma_4)\!-A_t m_0 \sigma_5 \big)
\sin2\beta\big]
(-y_3\! + \! y_2 \cos 2\beta +y_1 \sin 2\beta )
\nonumber\\
&-&\!\!\!
\big[ 2 m_{12} \sigma_1 \cos^2\beta\! -\!\frac{1}{2} \mu_0 \sigma_2 \sin 2\beta\!+\!
(2 m_{12}\sigma_4\! +\! A_t m_0 \sigma_5)\sin^2\beta \big] (y_1 \cos 2\beta\! -\!y_2 \sin 2\beta)
\Big]
\nonumber\\
&+&
(1/v^2)\,\Big[2\mu_0 \sigma_2 \cos 2\beta +\big[ 2 m_{12} (\sigma_1-\sigma_4) -A_t m_0 \sigma_5 \big]
\sin 2\beta\Big]
\Big[
-2 f^2 m_Z^2 (-\delta
\nonumber\\
&+&\!\! (4+\delta)\cos2\beta )\sin2\beta
\!+
4 v^2 (-y_3 +y_2\cos 2\beta +y_1 \sin 2\beta )(y_1 \cos 2\beta -y_2 \sin 2\beta)
\Big]
\Big\}
\eea
and
\bea
\Delta_{A_t^2}\!&\!\!=&\!\!\!\!\frac{-4 A_t}{ D} \Big\{ 8f^2 
(y_2\sin 2\beta-2y_1\cos2\beta) 
 \Big[ (m_0/2)\big(\mu_0\sigma_3\cos 2\beta+
(2A_tm_0\sigma_6 - m_{12}\sigma_5)
\sin 2\beta\big)   
 \nonumber\\
& \times&\!\!\!\! 
(-y_3+y_2\cos2\beta +y_1\sin 2\beta )
+m_0\sin\beta\big[\mu_0\sigma_3\cos\beta+
(-m_{12}\sigma_5+2A_t m_0\sigma_6)\sin\beta\big]  
\nonumber\\
&\times&\!\!\!\!(y_1\cos 2\beta-y_2\sin 2\beta)\Big]+(f^2/v^2) m_0\big[2\mu_0\sigma_3\cos 2\beta+
(-m_{12}\sigma_5+2A_tm_0\sigma_6) \sin 2\beta\big] 
\nonumber\\
&\times&\!\!\!\!\Big[-2f^2 m_Z^2
\big[-\delta+(4+\delta)\cos2\beta\big]\sin2\beta
+4v^2(-y_3+y_2\cos2\beta+y_1\sin2\beta)  
\nonumber\\
&\times&\!\!\!\!(y_1\cos2\beta-y_2\sin2\beta)\Big]- 
(4/v^2)m_0\sin\beta\Big[\big(f^2(\delta m_Z^2+4y_2)+2v^2y_2y_3\big)\cos2\beta
\nonumber\\
&-& \!\!\big[f^2m_Z^2(4+\delta)+2v^2(-y_1^2+y_2^2)\big]\cos4\beta+2y_1(4f^2+v^2y_3-4v^2y_2\cos2\beta)\sin2\beta\Big]
\nonumber\\
&\times&\!\!\!\!\Big[\mu_0\sigma_3\cos\beta
\,\big[ 2 f^2+v^2y_3-v^2\, (y_2\cos2\beta+y_1\sin2\beta)\big] 
+\!\!(m_{12}\sigma_5-2A_tm_0\sigma_6)\sin \beta
\nonumber\\
&\times&\!\!\! \big[-f^2-v^2y_3
+v^2(y_2\cos2\beta+y_1\sin2\beta)\big]\Big]\Big\}.
\eea

\medskip
\noindent
Finally
\medskip
\bea
\!\!\! \Delta_{B_0^2}\!&\!=&\!-\frac{8B_0 m_0 \mu_0\sigma_8}{D}
\Big\{ \frac{\sin 2\beta}{v^2}\Big[\big(f^2(\delta\, m_Z^2+4y_2)
+2v^2 y_2 y_3 \big)\cos2\beta-\big[(4+\delta)f^2m_Z^2  \nonumber\\
&+&\!\!2v^2(-y_1^2+y_2^2)\big]\cos4\beta
+ 2y_1 ( 4f^2+v^2 y_3- 4 v^2 y_2\cos2\beta)\sin2\beta\Big]
\big[-2f^2-v^2y_3 \nonumber\\
&+&\!\!v^2(y_2\cos2\beta+y_1 \sin2\beta)\big]+\frac{f^2}{v^2}
\cos2\beta\Big[-2f^2m_Z^2\big[-\delta+(4+\delta)\cos2\beta\big]\sin2\beta  \nonumber\\
&+&\!\!4v^2(-y_3+y_2\cos2\beta
+y_1\sin2\beta)(y_1\cos2\beta-y_2\sin2\beta)\Big]  
\nonumber\\
&-&\!\!2f^2(2y_1\cos2\beta-y_2\sin2\beta)(-y_3\cos2\beta
+y_2\cos4\beta+y_1\sin4\beta)\Big\}
\eea

\begin{figure}[t!]
\centering\def\baselinestretch{1.}
\begin{tabular}{cc}  %
\includegraphics[height=0.25\textheight,width=0.45\textwidth]{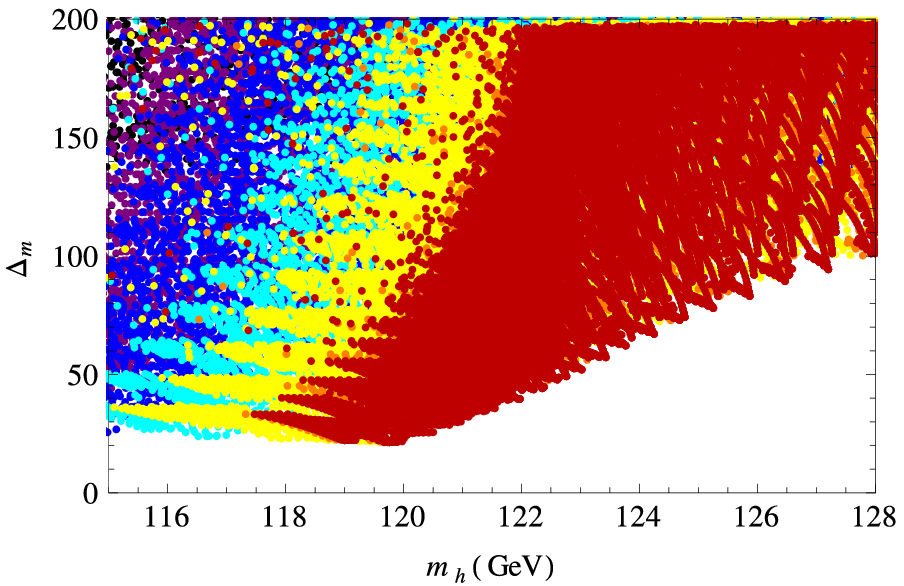}& 
\includegraphics[height=0.25\textheight,width=0.45\textwidth]{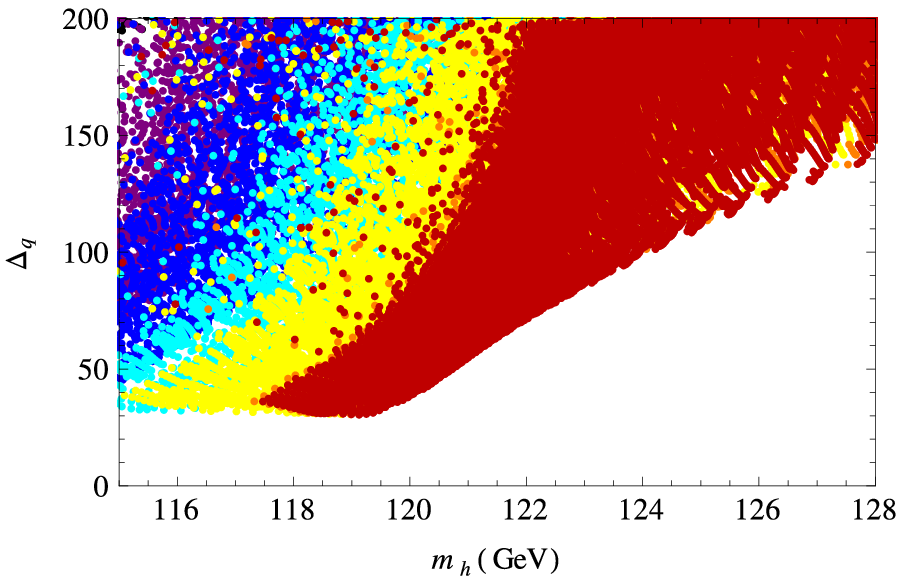}
\end{tabular}
\caption{{\protect\small The EW fine tuning $\Delta_m$ (left)
 and $\Delta_{q}$ (right) as  functions of the SM-like Higgs mass $m_h$ (in GeV), 
all evaluated at one-loop, for $\tan\beta\leq 10$. 
These plots have a fixed value  $\sqrt f=2.8$ TeV of the SUSY breaking scale and
$\tan\beta$  increases from left ($\tan\beta\leq 2.5$)
 to right ($\tan\beta=10$) as shown by  different colours:
black/leftmost region: $\tan\beta\leq 2.5$;
purple:  $2.5\!\leq\tan\beta\! \leq 4$;
blue: $4\!\leq\tan\beta \leq\! 4.5$;
cyan: $4.5\!\leq\tan\beta \leq\! 5.5$;
yellow: $5.5\leq\tan\beta \leq 9.5$;
red/rightmost region: $\tan\beta=10$ (a larger $\tan\beta$ region is on top of 
that of smaller $\tan\beta$).
For $m_h=126$ GeV,  
 minimal $\Delta_m\approx 80$ and $\Delta_q\approx 120$ while
in the corresponding constrained MSSM minimal values  (for $\tan\beta<55$),
 $\Delta_m\sim \Delta_q\approx 800-1000$, too large to be shown here;
 for details see figures 1-8 in \cite{KR}. The wide range of values for $m_h$ was 
chosen only to display the $\tan\beta$ dependence and to allow for the 2-3 GeV 
theoretical error of $m_h$ \cite{error}.}}
\label{higgsp2}
\end{figure}

\medskip\noindent
The denominator $D$ used in the above formulae is
\medskip
\bea
D&\equiv &
2 f^2 \Big[ \big[ f^2 (m_Z^2\delta +4 y_2 ) +2 v^2 y_2 y_3 \big] \cos 2\beta 
- 
\big[ (4+\delta) f^2 m_Z^2 +2 v^2 (y_2^2-y_1^2)\big]\cos4\beta
\nonumber\\
&+&\!\!
2 y_1 (4 f^2 +v^2 y_3 -4 v^2 y_2 \cos 2\beta )\sin 2\beta \Big]
\Big[
8 (m_Z^2/v^2) \big( \cos^2 2\beta +\delta \sin^4\beta \big)
+
(4/f^2)\, (-y_3
\nonumber\\
&+& \!\!
y_2\cos 2\beta +
 y_1\sin 2\beta \,)^2
\Big]
- 
(1/v^2) \,\Big[ 
-4 v^2  (-y_3+ y_2 \cos 2\beta +y_1 \sin 2\beta)(y_1 \cos 2\beta
\nonumber\\
& -&\!\!
y_2 \sin 2\beta )
+
f^2 m_Z^2 \big(-2 \delta \sin 2\beta +(4+\delta)\sin 4\beta\big)\Big]^2 
\eea

\medskip\noindent
In the above expressions  we introduced the notations:
\medskip
\bea
y_1 &\equiv& \mu_0 (m_{12} \sigma_2 + A_t \,m_0 \,\sigma_3 + B_0 m_0 \sigma_8)
\nonumber\\
y_2 &\equiv& - m_{12}^2 (\sigma_1-\sigma_4)- m_0 ( m_0 - A_t m_{12} \sigma_5 
+ A_t^2 m_0 \sigma_6 -m_0 \sigma_7)
\nonumber\\
y_3 &\equiv& y_2 + 2 \sigma_1 m_{12}^2 + 2 m_0^2,\,\,\,
y_4 \equiv A_t m_{12} \sigma_5 - 2 A_t^2 m_0 \sigma_6 +2 m_0 \sigma_7
\label{ys}
\eea

\medskip\noindent
The expressions for $\Delta_{\gamma^2}$ simplify considerably if
one turns off the quantum corrections to the soft terms 
($\sigma_{1,2,..6}=0$, $\sigma_{7,8}=1$).
We checked that  in the limit of large $f$, $\Delta_{\gamma^2}$
recover  the analytical results for fine tuning
at one-loop found in \cite{Cassel:2009ps} for the 
constrained MSSM (plus corrections $\cO(1/f^2)$). 
One also recovers from the above expressions for $\Delta_{\gamma^2}$ 
the results in eqs.(\ref{rr}).

\begin{figure}[t!]
\centering\def\baselinestretch{1.}
\begin{tabular}{cc} 
\includegraphics[height=0.25\textheight,width=0.45\textwidth]{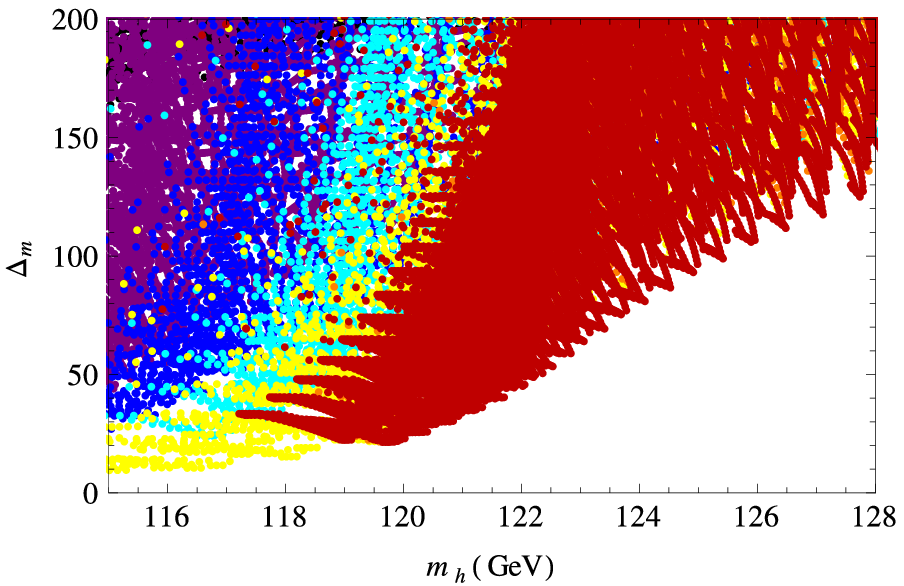}&
\includegraphics[height=0.25\textheight,width=0.45\textwidth]{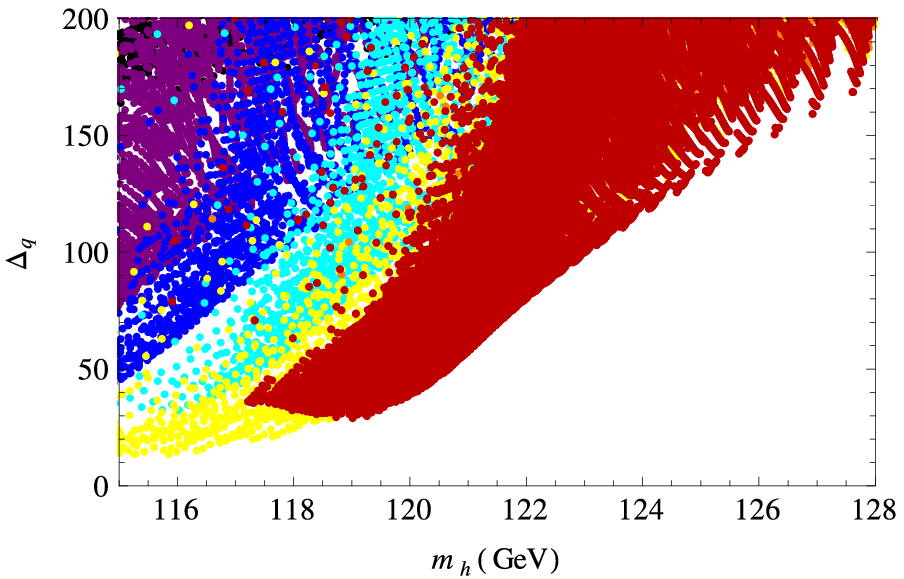}
\end{tabular}
\caption{{\protect\small 
$\Delta_m$ (left) and $\Delta_q$ (right), with similar considerations as 
for Figure~\ref{higgsp2} but with $\sqrt f=3.2$ TeV. In this case, minimal $\Delta_m=105$ and
$\Delta_q=145$ for $m_h=126$ GeV.
}}
\label{higgsp3}
\end{figure}

\begin{figure}[t!]
\centering\def\baselinestretch{1.}
\begin{tabular}{cc} 
\includegraphics[height=0.25\textheight,width=0.45\textwidth]{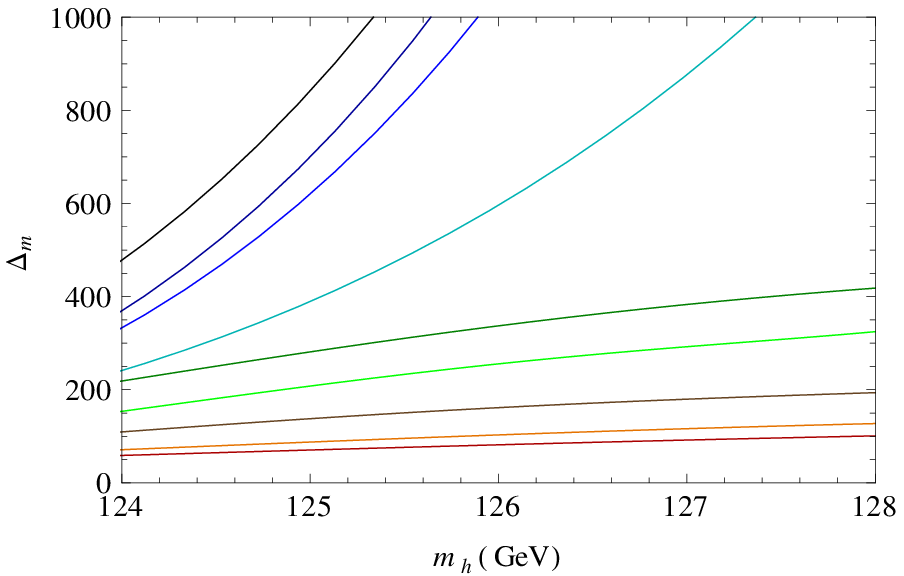}&
\includegraphics[height=0.25\textheight,width=0.45\textwidth]{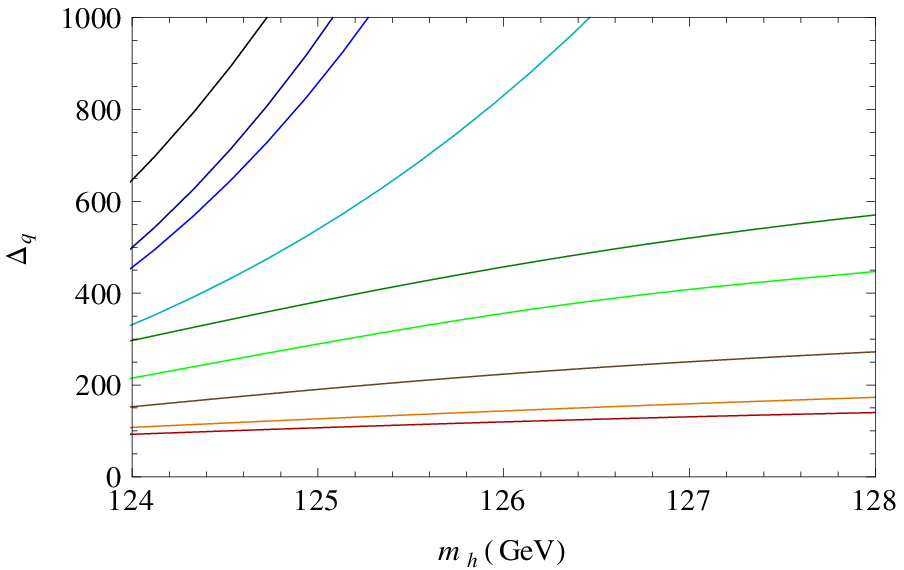}
\end{tabular}
\caption{{\protect\small
The dependence of {\it minimal} $\Delta_m$ (left) and $\Delta_q$ 
(right) on $m_h$ (GeV) for 
different $\sqrt f$, for fixed $\tan\beta=10$ with the other parameters
allowed to vary.  We allowed a $\pm 2$ GeV (theoretical) 
error for $m_h$ \cite{error} about the central value of $126$ GeV.
For a fixed $m_h$ the minimal values of  $\Delta_m$, $\Delta_q$ increase
as we increase $\sqrt f$ from the lowest to the top curve, in this order: 
$2.8$ TeV (the lower/red curve),  $3.2$ TeV (orange),
$3.9$ TeV (brown), $5$ TeV (green), $5.5$ TeV (dark green),
$6.3$ TeV (cyan), $7.4$ TeV (blue),  
$8$ TeV (dark blue), $8.7$ TeV (black/top curve).
The lowest two curves (red, orange) 
 correspond to the  minimal values of $\Delta_m$ and $\Delta_q$ in Figures~\ref{higgsp2},~\ref{higgsp3}. 
For large enough $\sqrt f\geq 10$ TeV, one recovers the MSSM-like values of $\Delta_m$, $\Delta_q$ for 
a similar $m_h$.}}
\label{lines}
\end{figure}

\subsection{Numerical results}

Using the results in eqs.(\ref{dmu}) to (\ref{ys})  we evaluated 
$\Delta_m$ and $\Delta_q$ for fixed values of the SUSY  breaking scale in the hidden
sector $\sqrt f$ for $\tan\beta \leq 10$, subject to the EW constraints (for a discussion of
these, see  \cite{KR}). Note that imposing the higgs mass range of $126\pm (2\, {\rm to}\, 3)$ 
GeV (to allow for the theoretical error \cite{error}) automatically
respects these constraints \cite{KR}.
For a rapid convergence of the perturbative expansion 
in $1/f$ of the Lagrangian we demanded that $m_{soft}^2/f<1/4$. 
The results are shown in Figures~\ref{higgsp2}, \ref{higgsp3}, \ref{lines}.

For  $m_h=126$ GeV we find {\it minimal} values of
$\Delta_m\approx 80$ and $\Delta_q\approx 120$ for $\sqrt f=2.8$ TeV (Figure~\ref{higgsp2})
and
$\Delta_m\approx 105$ and $\Delta_q\approx 145$ for $\sqrt f=3.2$ TeV (Figure~\ref{higgsp3}).
These values of $\sqrt f$ are well above the current lower bound of $\approx 700$ GeV
 \cite{nonlinearMSSM,Brignole:2003cm,lowerbound,Antoniadis:2004se}.
As one increases $\tan\beta$ for a given $m_h$, $\Delta_m$ or $\Delta_q$ decreases, 
as shown by the colour encoding corresponding to fixed $\tan\beta$
 in Figures~\ref{higgsp2},~\ref{higgsp3}; this is also valid in the  MSSM as seen in 
Figures~3, 4, 5 in the first reference in \cite{Cassel:2010px}.
These values for fine tuning are already ``acceptable'' and
 significantly below the {\it minimal} values  in the constrained MSSM where
for  $m_h\approx 126$ GeV, $\Delta_{m,q}\approx 800-1000$, see Figures~1-8 in \cite{KR},
obtained after scanning over all $2\leq \tan\beta\leq 55$.

The reduced values  of $\Delta_m$ and $\Delta_q$ are due to the fact that $m_h$ is significantly
above that of the constrained MSSM already at the classical level, see eqs.(\ref{mhH}) to (\ref{lim})
for $\delta=0$, where values of $120-126$ GeV are easily achieved, so only very 
small quantum corrections are actually needed (unlike in the MSSM).
This is a consequence of the (classically)
 increased effective quartic  higgs coupling.
Also notice that minimal values of $\Delta_m$ and $\Delta_q$ have a 
similar dependence on $m_h$ and are only mildly different in size,
as also noticed for the MSSM \cite{KR}.

In Figure~\ref{lines} we presented the minimal values of $\Delta_m$ and $\Delta_q$
as functions of $m_h$ for fixed $\tan\beta=10$ for different values of the SUSY breaking scale 
from $\sqrt f=2.8$ TeV to  $8.7$ TeV. When increasing
$\sqrt f$ to larger values, in the region above $10$ TeV, 
the effects of the additional quartic terms in the scalar 
higgs potential are rapidly suppressed and one recovers the usual constrained MSSM-like scenario
with similar UV boundary conditions, with larger fine tuning
for the same $m_h$ and with  minimal $\Delta_{q,m}\!\sim\! \exp(m_h/{\rm GeV})$ (see the
top curves in Figure~\ref{lines}). This exponential behaviour is characteristic to MSSM-like models
due to (large) quantum corrections to the Higgs mass \cite{Cassel:2010px}.
Relaxing the UV universality boundary condition for the gaugino masses reduces $\Delta_{m,q}$ further, 
similar to the MSSM \cite{KR,King,Horton:2009ed},  
by a factor of $\approx 2$ from the values given by the curves in Figure~\ref{lines}. 
Thus, values of $\sqrt f$ of up to 5-6 TeV
can still give an EW fine tuning of about $\sim 100$, for the low $\tan\beta$ regime considered here.

The case of constrained ``non-linear'' MSSM at small $\tan\beta\leq 10$, for which  
we found  ``acceptable'' values for $\Delta_{m,q}$, is the most conservative scenario.
We saw in Figures~\ref{higgsp2},\ref{higgsp3}
that for the same $m_h$ a larger $\tan\beta$ reduces fine tuning 
and this behaviour  continues to $\tan\beta\sim 40-50$.
Then additional Yukawa couplings also play a significant role at larger $\tan\beta$
and reduce fine-tuning further by improving the radiative EW symmetry breaking
for the same $m_h$ (this is because radiative EW symmetry breaking effects are enhanced
relative to opposite, QCD ones that increase fine-tuning \cite{Cassel:2010px}).
We thus expect that for the case of large $\tan\beta$ with additional Yukawa couplings included
the values quoted here for $\Delta_m$, $\Delta_q$ be maintained or reduced further.

\section{Conclusions}

The significant amount of EW fine tuning $\Delta$ present in the MSSM-like
models for $m_h\approx 126$ GeV  has prompted an increased
interest in finding ways to reduce its value. 
This is motivated by the fact that $\Delta$  is usually regarded as a
measure of the success of SUSY in solving the hierarchy problem.
Additional reasons to seek a low $\Delta$ exist, 
from the relation of the EW fine tuning to the
 variation   $\delta\chi^2$ about the minimal chi-square $\chi^2_{min}$
and the s-standard deviation upper bound on $\delta\chi^2$ usually sought
in the data fits. 
Reducing $\Delta$ can indeed be  achieved, but it usually requires the 
introduction of additional fields  in the visible sector, 
beyond those of the original model. For example one can consider MSSM-like
 models with additional, massive gauge singlets present, extra gauge symmetries, etc.

Another view is that a large EW fine tuning may indicate a problem with our
understanding of supersymmetry breaking.  Motivated by this we considered
the case of MSSM-like models with a low scale of supersymmetry breaking in the hidden
sector, $\sqrt f\sim$ few TeV.
 As a result of this, sizeable quartic effective interactions are present in the 
Higgs potential, generated by the exchange of the auxiliary  field of the goldstino superfield.
Such couplings are proportional to the ratio of the soft breaking terms $m_{soft}$ in the visible sector 
to the SUSY breaking scale $\sqrt f$ of the hidden sector. Thus,
such couplings are significant in 
 models with  $\sqrt f \sim$ few TeV and are  negligible
 when $\sqrt f$ is large,   which is the usual MSSM scenario. 
These couplings have significant implications for the higgs mass and the EW fine-tuning.
This behaviour is generic in low-scale SUSY models.

For the most conservative case of a constrained  ``non-linear'' MSSM  model and   
at low $\tan\beta$, we computed the level of EW scale fine tuning 
measured by two definitions for $\Delta$ ($\Delta_m$, $\Delta_q$). We examined  $\Delta_{m,q}$
  as a function of the SM-like  higgs mass,  in the one-loop  approximation for these quantities.  
The results show that for $m_h\approx 126$ GeV, fine tuning
is  reduced  from  {\it minimal} values of 
$\approx 800-1000$ in the constrained MSSM  to more acceptable  values of 
$\sim 80-100$ in our model with  $\sqrt f\sim 2.8 - 3.2$ TeV. These values for $\Delta$
 are expected to be further reduced by considering non-universal gaugino masses. We argued that a 
similar reduction of $\Delta$ is expected at large $\tan\beta$ in our model.
For larger $\sqrt f$, usually above $10$ TeV, one recovers the case of MSSM-like  
models. Unlike other similar studies, this reduction was possible  without additional 
fields in the visible sector and depends only on the ratio(s) $m_{soft}^2/f$;
one may even consider increasing both $m_{soft}$ and $\sqrt f$, while keeping
 their ratio fixed.

We assumed that in our case the sgoldstino  was massive enough and
integrated out, by using the superfield constraint that decouples it from the low energy.
Corrections to our result can then arise from the scalar potential for the sgoldstino
that depends on the structure of its Kahler potential (that gives mass to it) and superpotential in 
the hidden sector. 
Another correction can arise from  future experimental constraints that 
may increase the lower  bounds on the value of   $\sqrt f$, currently near $\approx700$ GeV, 
if no supersymmetry or other new physics signal is found.

\bigskip\medskip
\noindent
{\bf Acknowledgements: }
This work was supported in part by the European Commission under the ERC 
Advanced Grant 226371.
The work of D.~M.~Ghilencea  was supported by  a grant of the 
Romanian Research Council   project number PN-II-ID-PCE-2011-3-0607  and in part by 
National Programme  `Nucleu'  PN 09 37 01 02.

\section*{Appendix}

\def\theequation{A-\arabic{equation}}
\def\thesubsection{A}
\setcounter{equation}{0}
\def\thefigure{A-\arabic{figure}}

The coefficients $\sigma_i$ at the EW scale, used in the text,
 eq.(\ref{ms}) have the expressions:
\medskip
\bea\label{coefficients}
\sigma_1(t_z)&=&0.532,  \qquad \qquad\qquad \qquad\qquad \qquad
\sigma_2(t_z)=0.282\,(4.127\,h_t^2-2.783)(1.310-h_t^2)^{1/4}\nonumber\\
\sigma_3(t_z)&=& -0.501\,h_t^2\,(1.310-h_t^2)^{1/4}, \qquad \quad\,\,
\si_4(t_z)=0.532-5.233\,h_t^2+1.569\,h_t^4\nonumber\\
\si_5(t_z)&=& 0.125\,h_t^2\,(10.852\,h_t^2-14.221), \qquad 
\si_6(t_z)= -0.027\,h_t^2\,(10.852\,h_t^2-14.221)\nonumber\\
\si_7(t_z)&=&1-1.145 \,h_t^2,  \qquad \qquad\qquad \qquad\quad\,
\si_8(t_z)= 1.314 \,(1.310-h_t^2)^{1/4}
\eea

\medskip\noindent
where $h_t$ is evaluated at $m_Z$ and
$m_t=h_t(t_{m_t})\,(v/\sqrt 2)\sin\beta$, ($v=246$ GeV), $t=\ln \Lambda^2/q^2$, 
$t_z=\ln \Lambda^2_{UV}/m_Z^2$.

\end{document}